\documentclass{JHEP}

\newtheorem{prop}{Proposition}
\newtheorem{lemma}[prop]{Lemma}

\usepackage{epsfig,amssymb,amsmath}

\newcommand{\rom}[1]{\mathrm{#1}}

\usepackage{graphicx,subfigure}
\usepackage{epstopdf}
\usepackage{epsfig,amssymb,amsmath}
\usepackage{epsfig}
\usepackage{graphicx}
\usepackage{enumerate}
\usepackage{amsmath}

  \newcommand{\be}{\begin{equation}}
  \newcommand{\ee}{\end{equation}}
  \newcommand{\bea}{\begin{eqnarray}}
  \newcommand{\eea}{\end{eqnarray}}
  
    \newcommand{\eps}{\epsilon}

  \newcommand{\s}{\sigma}
  
  \newcommand{\nn}{\nonumber}

\newcommand{\sech}{\,\text{sech}\,}

\newcommand{\curl}{\mbox{curl}\:}

\def\cM{\mathcal{M}}

\newcommand{\beq}{\begin{equation}}
\newcommand{\eeq}{\end{equation}}
\newcommand{\beqs}{\begin{eqnarray}}
\newcommand{\eeqs}{\end{eqnarray}}

\newcommand{\DD}{{\mathcal{D}}}

\preprint{ULB-TH/11-15}

\title{Asymptotic Flatness, Taub-NUT, and Variational Principle}

\author{Amitabh Virmani \\
 Physique Th\'eorique et Math\'ematique,  \\ Universit\'e Libre de
     Bruxelles and International Solvay Institutes, Bruxelles,  Belgium,  \texttt{avirmani@ulb.ac.be}}

\abstract{Using hyperbolic temporal and spatial cut-offs to define 4d asymptotically flat spacetimes, we show that supertranslation ambiguities in the asymptotic fields can all be removed even in the presence of gravitational magnetic charges. We then show that configurations with different electric and magnetic four-momenta can be consistently considered in the Mann-Marolf variational principle. This generalizes the previous result where
 variations over asymptotically flat configurations with 
fixed magnetic four-momenta were considered. We also express Kerr Taub-NUT metric to the leading and next to leading order in the Beig-Schmidt form, and compare the asymptotic form with the tensor harmonics on 3d de Sitter space.
}

\date{June 2011}


\begin{document}

\section{Introduction}
In the light of recent developments it is tempting to speculate that gravitational electric-magnetic duality
with certain appropriately defined asymptotically flat boundary conditions can be realized in some precise sense.

A first set of clues in this direction come from the generalized geometry of M-theory series of papers  \cite{Berman:2010is, Berman:2011pe, Hillmann:2009ci} where the nature of SL(5), SO(5,5), and E7(7) U-dualities of M-theory has been explored.  This approach when extended to the E8(8) symmetry of M-theory naturally incorporates gravitational electric-magnetic duality; see \cite{Nicolai1, Nicolai2, Nicolai3} for the relevant previous work. This is because the Ehlers SL(2) of vacuum four-dimensional gravity is a subgroup of the M-theory E8(8). The compact SO(2) generator  of the Ehlers SL(2) is expected to be the generator of the gravitational electric-magnetic duality.

A second set of clues come from the fact that the idea has been successfully realized for linearized
Einstein gravity around flat space.  The spin-2 Fronsdal action is invariant under
an SO(2) rotation \cite{Nieto:1999pn, Henneaux:2004jw, Barnich:2008ts}. See also \cite{Argurio:2008zt, Argurio:2008zt2, Henneaux:2007ej, Hull:2001iu} and references therein. However, it is not clear how the considerations of \cite{Nieto:1999pn, Henneaux:2004jw, Barnich:2008ts} are to be extended to the non-linear theory: it has been shown by Deser and Seminara \cite{Deser:2005sz} that
this duality cannot be extended perturbatively to the 3-vertex in Einstein
gravity using a proof similar to the one showing that electric-magnetic duality of free Maxwell theory cannot be extended to Yang-Mills theory \cite{Deser:1976iy}. Given this result, gravitational duality, if correct, can only be realized in certain non-perturbative or non-local sense not amenable to the treatment of \cite{Deser:2005sz, Deser:1976iy}, e.g., perhaps complying with the ideas put forward in \cite{Berman:2010is, Berman:2011pe, Hillmann:2009ci}.

In an attempt to make some of this discussion more precise we explore a notion of asymptotic flatness at spatial infinity where we allow for the presence of gravitational magnetic charges. The case of cylindrical cut-offs to define asymptotically flat spacetimes has been previously treated in the literature \cite{Bunster:2006rt}. In this paper, we work instead with hyperbolic temporal and spatial cut-offs to define 4d asymptotically flat spacetimes, and in particular use the Beig-Schmidt expansion \cite{BS, B}. We show that supertranslation and logarithmic translation ambiguities in the asymptotic fields can all be removed even when non-zero gravitational magnetic charges are present. In the absence of gravitational magnetic charges this is a well known result \cite{Ashtekar:1978zz, AshRev, BS, B, ABR, Ashtekar:1991vb}. However, to the best of of our knowledge a complete argument has not been presented before when non-zero gravitational magnetic charges are present. Some comments already appear in \cite{AM2, Ashtekar:1991vb} and more recently in \cite{CDV}. We fill this gap here.

We then show that configurations with different gravitational magnetic charges can be consistently varied in the Mann-Marolf \cite{MM} variational principle, i.e., competing histories can have different electric as well as magnetic four-momenta. This generalizes the previous result of \cite{MM} where variations over asymptotically flat configurations with only
fixed magnetic four-momentum were considered. This part of our investigation is motivated by the observation of \cite{Bunster:2006rt} that the Taub-NUT metric is asymptotically flat at spatial infinity in the sense of Regge and Teitelboim \cite{Regge:1974zd} including their parity conditions. From this observation reference \cite{Bunster:2006rt} inferred that one can consistently consider configurations with different electric and magnetic masses in the Regge-Teitelboim variational principle. We establish this same result in the covariant formalism of \cite{MM}. In showing this we perform integrations by parts which are not completely justified. This is due to the presence of certain singularities in the asymptotic metric. The investigation in that aspect should be regarded as preliminary. The main point we want to emphasize, as reference \cite{Bunster:2006rt} did for the Regge-Teitelboim variational principle, is that Taub-NUT metric satisfies all boundary conditions required to be considered in the action principle of \cite{MM}. Geometrical difficulties as pointed out by Misner \cite{Misner} remain and are not addressed in this work. Other studies of asymptotic properties of Taub-NUT solution include \cite{RS, AS} at null infinity and \cite{Argurio:2008zt2, BNS} at spatial infinity.

The rest of the paper is organized as follows. In section \ref{prelims} we introduce our notion of asymptotic flatness. In section \ref{sec:MM} we present the Mann-Marolf variational principle and show that the on-shell action is finite for configurations with non-zero gravitational magnetic charges. We also calculate the on-shell value of the action. We then show that configurations with non-zero gravitational magnetic charges can be varied in the action principle. In section \ref{app:KerrNut} we express Kerr Taub-NUT metric to the leading and the next to leading order in
the Beig-Schmidt form, and compare the asymptotic form with tensor harmonics on three-dimensional de Sitter space. We conclude in section \ref{conclu} with a brief summary and comments on future directions. Certain calculational details are relegated to the two Appendices.


\section{Preliminaries}
\label{prelims}
In this section we introduce our notion of asymptotic flatness and characterize the asymptotic form of the metric at leading order. Our notion of asymptotic flatness is based on the Beig-Schmidt expansion. We fix boundary conditions to obtain the Poincar\'e group as the asymptotic symmetry group. The new element in the following discussion is that we allow for certain singular terms in the metric expansion to deal with gravitational magnetic (Taub-NUT) charges. We exclusively work with a coordinate based definition.

\subsection{Beig-Schmidt Expansion}

In the large $\rho$ expansion the metric takes the form \cite{BS, B}\footnote{See the companion paper \cite{CDV} for notational details. We exclusively use the notation of conventions of this paper. We use $\DD_a$ to denote the torsion-free covariant derivative on the unit hyperboloid. The radial coordinate $\rho$ is associated to some asymptotically Minskowski coordinates $x^\mu$ via $\rho^2 = \eta_{\mu \nu} x^\mu x^\nu$.}
\bea
ds^2 = g_{\mu \nu }dx^{\mu}dx^{\nu} & = &\left(1 + \frac{\s}{\rho} \right)^2 d \rho^2 + h_{ab}dx^a dx^b \nn \\
& =& \left(1 + \frac{\s}{\rho} \right)^2 d \rho^2 + \rho^2 \left( h^{(0)}_{ab} + \frac{h^{(1)}_{ab}}{\rho} + \frac{h^{(2)}_{ab}}{\rho^2} + \mathcal{O}(\rho^{-3})\right), \label{metric}
\eea
where $h^{(0)}_{ab}dx^a dx^b$ is the metric on the unit hyperboloid or, equivalently, on three-dimensional de Sitter space $dS_3$,
\be
ds^2_{(0)} \equiv h^{(0)}_{ab}dx^a dx^b = -d\tau^2 + \cosh^2\tau (d\theta^2 + \sin^2\theta d\phi^2).\label{h0}
\ee


The field $\sigma$, usually referred to as the mass aspect, is assumed to be a smooth function on the unit hyperboloid.  We consider fields $h^{(1)}_{ab},h^{(2)}_{ab}$, etc, admitting certain singularities on the hyperbolic space. These singularities can be characterized in several ways, using either Dirac-Misner strings or using two patches on the hyperbolic space and considering non-trivial transition functions between the patches in complete analogy with the treatment of magnetic monopoles in electromagnetism. We consider the formulation using two patches. We expect that all results can be reformulated using Dirac-Misner strings as well. We denote as the northern semi-hyperbolic space $\mathcal N$ and the southern semi-hyperbolic space $\mathcal S$, for both of which the spacetime metric is \eqref{h0} with $\tau \in \mathbb R$, $\phi \in [0,2\pi)$ and, respectively, $\theta \in (0,\pi/2]$ and $\theta \in [\pi/2,\pi)$. The hyperboloid is covered by these two patches intersecting at the equator $\theta = \pi/2$.

The singularities in the field $h^{(1)}_{ab}$  are characterized as follows\footnote{Singularities in $h^{(2)}_{ab}$ and other higher order fields are introduced through the asymptotic expansion, as nonlinear terms constructed using $h^{(1)}_{ab}$ act as sources for asymptotic equations for $h^{(2)}_{ab}$ and other fields.}. We allow $h^{(1)}_{ab}$ to have non-trivial transition functions between the northern and southern patches, requiring that the four-dimensional Riemann tensor be continuous across the patches. As will be described shortly, the most general form of the transition function can then be written as
\be
h^{(1)}_{ab}\big{|}_{\mathcal S} - h^{(1)}_{ab}\big{|}_{\mathcal N} = \sum_{\mu=0}^3 N_{(\mu)} \delta k_{(\mu)}{}_{ab} \label{diffh1}.
\ee
The numbers $N_{(\mu)}$ encode the NUT charges.

The relaxation of the regularity requirement on $h^{(1)}_{ab}$ constitute an increase in the set of allowed configurations considered with respect to the usual notions of asymptotic flatness at spatial infinity. Let us therefore now look at the changes (if any) that our relaxed boundary conditions bring in to the discussion of allowed asymptotic symmetries.  The set of diffeomorphisms  preserving the Beig-Schmidt form \eqref{metric}  are well understood,
see \cite{CDV} for a concise review. In an asymptotically cartesian coordinate system with $\rho^2 = \eta_{\mu \nu} x^{\mu} x^{\nu}$, diffeomorphisms
$
\bar{x}^\mu = L_{\nu}^{\mu}x^\nu + T^\mu + S^\mu(x^a)  + o(\rho^0), \label{diffeos}
$ preserve the form of the metric \eqref{metric}. The transformations generated by the constants $L_{\nu}^{\mu}$ and $T^\mu$ form the Poincar\'e group, while transformations generated by angle dependent translations $S^\mu(x^a)$ are the so-called supertranslations.

Since supertranslations depend arbitrarily on the angular coordinates $x^a$, in particular on the time coordinate $\tau$,  the associated charges are in general not conserved. In fact, the approach that a large body of work on asymptotic flatness at spatial infinity has taken is to strengthen the boundary conditions so that the freedom of performing supertranslations is eliminated. See, however, \cite{CD} for an alternative point of view where a set of supertranslations satisfying $(\Box + 3) \omega =0$ are allowed to act on the phase space. Indeed, supertranslation ambiguities can be removed \cite{Ashtekar:1978zz, B, Ashtekar:1991vb, ABR}, for instance, by demanding the leading order asymptotic Weyl curvature to be purely electric.
This condition can be fullfilled by choosing $h^{(1)}_{ab} = - 2 \sigma h^{(0)}_{ab}$. This condition removes altogether the possibility of NUT charges. Since our crucial point is to allow for NUT charges, clearly we will not demand the leading order asymptotic Weyl curvature to be purely electric. Non-zero asymptotic magnetic Weyl curvature brings in several new elements to the discussion. As will be explained in detail in the next subsection, certain forms of the non-zero leading order magnetic part of the Weyl tensor leads to singularities in the boundary metric $h_{ab}$, in particular in $h^{(1)}_{ab}$. As a result $h^{(1)}_{ab}$ can no longer be restricted to the form $h^{(1)}_{ab} = - 2 \s h^{(0)}_{ab}$. Thus, instead of requiring $h^{(1)}_{ab} = - 2 \sigma h^{(0)}_{ab}$, we require
\be
h^{(1)}_{ab} = - 2 \sigma h^{(0)}_{ab} + k_{ab},
\ee
where the form of $k_{ab}$ is fixed to remove supertranslation ambiguities. This can be done in numerous ways: for example, since $S^\mu(x^a)$ depends \emph{smoothly} on angular coordinates, requiring $k_{ab}$ to contain only singular terms completely fixes the supertranslation  ambiguity in exactly the same way as demanding $h^{(1)}_{ab} = - 2 \sigma h^{(0)}_{ab}$ fixes it when the leading order asymptotic Weyl curvature is purely electric. We present the precise arguments in the next subsection.

In addition, to remove ambiguities related to the so-called logarithmic translations, see \cite{CDV} for a recent discussion, we continue to impose that $\sigma$ in the harmonic decomposition on $dS_3$ does not contain the four lowest parity odd harmonics. We recall from that the four lowest parity odd harmonics are solutions to the equation $\DD_a \DD_b H + H h^{(0)}_{ab} = 0$. Four independent solutions $\zeta_{(\mu)}$, $\mu=0,\ldots3$ for this equation are
\bea
\zeta_{(0)} &=& \sinh\tau, \nn \\ 
\zeta_{(1)} &=& \cosh\tau \cos\theta,\nn \\
\zeta_{(2)} &=& \cosh\tau \sin\theta\cos\phi,\nn \\ 
\zeta_{(3)} &=& \cosh\tau \sin\theta \sin\phi. \label{zetas}
\eea
The vectors $\DD_a \zeta_{(\mu)}$ are four conformal Killing vectors on $dS_3$. These conformal Killing vectors precisely correspond to four translations in the asymptotically flat context.
\subsection{First Order Structure}
\label{1st}
In this section we analyze the leading  order structures in the Beig-Schmidt expansion. The first order equations of motion are studied and the supertranslation fixed asymptotic form of metric $h^{(1)}_{ab}$ admitting NUT charges is presented.

It is convenient to introduce the variable
\be
k_{ab} = h^{(1)}_{ab} +2 \sigma h^{(0)}_{ab}\label{def:kab},
\ee
which is in one-to-one correspondence with the first-order three-dimensional metric $h^{(1)}_{ab}$. The equations of motion at first order in the Beig-Schmidt expansion read as follows \cite{BS} \bea
\text{(i)}&&\square \sigma + 3 \sigma = 0,\\
\text{(ii)}&& \DD^b k_{ab} - \DD_a k = 0,\\
\text{(iii)}&& (\square -3)k_{ab} +k h^{(0)}_{ab} -\DD_a \DD_b k =0,
\eea
where $\square = \DD^a \DD_a$ and $k = k_{a}{}^a$. One can rewrite these equations
in a more enlightening form upon introducing the following two tensors derived from the following two lemmas:
\begin{lemma} On the three dimensional hyperboloid, a scalar satisfying $\square \Phi + 3\Phi = 0$ defines a symmetric tensor
\begin{equation}
T_{ab} = \DD_a \DD_b \Phi + h^{(0)}_{ab}\Phi,
\end{equation}
obeying $(\curl T)_{ab} := \eps_a{}^{cd} \DD_{c}T_{db} = 0$, $T^a_{a}=0$, $\DD^b T_{ab}=0$ and $\square T_{ab}-3T_{ab}=0$.
\end{lemma}
\begin{lemma}
\label{lemma:2}
On the three dimensional hyperboloid, any symmetric tensor $P_{ab}$ satisfying $\DD^b P_{ab} - \DD_a P^b_{\;\; b} = 0$, $(\square -3)P_{ab} +P^c_{\;\; c}h^{(0)}_{ab} -\DD_a \DD_b P^c_{\;\; c} =0$ defines a  tensor
\begin{equation}
T_{ab} = \epsilon_a^{\;\; c d} \DD_c P_{d b},
\end{equation}
obeying $T^a_{a}=0$, $\DD^b T_{ab}=0$, $\square T_{ab}-3T_{ab}=0$, and $(\curl T)_{ab} =0$.
\end{lemma}

These two lemmas can be proven with a little bit of algebra on 3d maximally symmetric spaces. The required properties of tensors on $dS_3$ are listed in the beginning of Appendix A of \cite{CDV}. From the scalar $\sigma$ and the tensor $k_{ab}$, we then define
\bea
E^{(1)}_{ab} = -\DD_a \DD_b \sigma - h^{(0)}_{ab} \sigma \, , &&  B^{(1)}_{ab} = \frac{1}{2}\epsilon_a^{\;\; c d} \DD_c k_{d b} \, .\label{dBk}
\eea
By construction, these two tensors enjoy the following properties
\bea
h^{(0)ab} B^{(1)}_{ab} = 0, \qquad E^{(1)}_{[ab]}  = 0, \qquad \DD_{[c} E^{(1)}_{a] b} = 0.
\eea
The equations of motion then become
\bea
\text{(i)}&&h^{(0)ab} E^{(1)}_{ab} = 0\,\nonumber ,\\
\text{(ii)}&& B^{(1)}_{[ab]}  = 0\, \label{EOM1},\\
\text{(iii)}&& \DD_{[c} B^{(1)}_{a] b} = 0.\nonumber
\eea
Therefore, on-shell, both $B^{(1)}_{ab}$ and $E^{(1)}_{ab}$ are symmetric, traceless, and curl-free. One can easily  check that these properties also imply that both tensors are divergence-free and are annihilated by the operator $(\square -3)$. In fact, tensors  $B^{(1)}_{ab}$ and $E^{(1)}_{ab}$ can be recognized as the first-order magnetic and electric part of the Weyl tensor, respectively.  Since these tensors are constructed from the curvature, they do not contain Dirac-Misner strings by construction. The tensor potential $k_{ab}$, however, might contain Dirac-Misner strings. We now turn to the classification of these singularities.
One can first prove the following lemma\footnote{A proof of this lemma was not given in \cite{Ashtekar:1978zz, AshRev}. A proof using harmonic decomposition of tensors on three dimensional de Sitter space has been recently presented in Appendix A of \cite{CDV}. See also appendix C of \cite{AM2}.} due to Ashtekar and Hansen \cite{Ashtekar:1978zz, AshRev}:
\begin{lemma}[Ashtekar-Hansen]\label{Ash} On the three-dimensional hyperboloid, any traceless curl-free divergence-free symmetric tensor $T_{ab}$ such that $\square T_{ab}= 3 T_{ab}$ can be written as
\begin{equation}
T_{ab} = \DD_a \DD_b \Phi + h^{(0)}_{ab}\Phi,
\end{equation}
with $\square \Phi + 3\Phi = 0$. The scalar $\Phi$ is determined up to the ambiguity of adding a combination of the four functions $\zeta_{(\mu)}$ \eqref{zetas}.
\end{lemma}
Using this lemma and equations of motion it immediately follows that both the first order electric and magnetic parts of the Weyl tensor admit scalar potentials. The scalar potential for $E^{(1)}_{ab}$ is exactly $(-\sigma)$. The scalar potential to $B^{(1)}_{ab}$ is a derived quantity that we denote as $\sigma^D$ (sigma-dual),
\begin{equation}
B^{(1)}_{ab} = \DD_a \DD_b \sigma^D + h^{(0)}_{ab}\sigma^D.
\end{equation}

As a result of the previous lemma, the physical content of these tensors can be further specified by solving explicitly the hyperbolic equation $\square \Phi +3\Phi =0$. The general solution is the sum of a function of $\tau$ times a spherical harmonic on the unit two-sphere,  $f_{lm}(\tau)Y_{lm}(\theta,\phi)$, $l = 0,1,\dots$, $\infty$, $m=-l,\dots,+l$,  where two independent solutions for $f_{lm}(\tau)$ exist since the hyperbolic equation is second order in time. We denote these solutions as $f_{lm}(\tau)$ and $\hat f_{lm}(\tau)$. One can then write
\be
\Phi = \sum_{\mu=0}^3 \left( \alpha_{(\mu)}  \zeta_{(\mu)}+ \hat \alpha_{(\mu)} \hat \zeta_{(\mu)}  \right) + \sum_{l \geq 2} \sum_{m=-l}^l \left( a_{lm} f_{lm}(\tau) +\hat a_{lm} \hat f_{lm}(\tau) \right) Y_{lm}(\theta,\phi),\label{decompphi}
\ee
where the first set of lowest harmonics are given in \eqref{zetas} and the other set of lowest harmonics are
\begin{eqnarray}
\hat \zeta_{(0)} &=& \frac{\cosh 2\tau}{\cosh \tau}, \nn \\
\hat \zeta_{(1)} &=& \left(2\sinh \tau+\frac{\tanh \tau}{\cosh \tau}\right)\cos\theta,\nn\\
\hat \zeta_{(2)} &=& \left(2\sinh \tau+\frac{\tanh \tau}{\cosh \tau}\right)\sin\theta\cos\phi,\nn\\
\hat \zeta_{(3)} &=& \left(2\sinh \tau+\frac{\tanh \tau}{\cosh \tau}\right)\sin\theta\sin\phi. \label{zetahat}
\end{eqnarray}
With these preliminaries, we are now in position to formulate the following important lemma\footnote{Again, a complete proof of this lemma was not given in \cite{BS}. A proof using harmonic decomposition of tensors on three dimensional de Sitter space has been recently presented in Appendix A of \cite{CDV}.} due to Beig and Schmidt \cite{BS}.
\begin{lemma}[Beig-Schmidt] \label{BS} On the three dimensional hyperboloid, any scalar $\Phi$ satisfying $\square \Phi + 3\Phi = 0$ and such that it does not contain the four lowest hyperbolic harmonics \eqref{zetahat} defines a symmetric, traceless, curl-free and divergence-free tensor $T_{ab} = \DD_a \DD_b \Phi + h^{(0)}_{ab}\Phi$ that can be written as
\begin{equation}
T_{ab} = \epsilon_{a}^{\;\, cd}\DD_c P_{d b},
\end{equation}
where $P_{ab}$ is a symmetric, traceless, and divergence-free regular tensor. This tensor is defined up to the ambiguity $P_{ab} \rightarrow P_{ab} + \DD_a \DD_b \omega + h^{(0)}_{ab}\omega$ where $\omega$ is an arbitrary scalar obeying $\square \omega +3\omega =0$.
\end{lemma}

This lemma shows that both $B^{(1)}_{ab}$ and $E^{(1)}_{ab}$ admit regular tensor potentials provided that they do not possess the four harmonics $\hat \zeta_{(\mu)}$ \eqref{zetahat} in their scalar potentials. The four harmonics $\hat \zeta_{(\mu)}$ in $B^{(1)}_{ab}$ exactly represent the singular terms for $k_{ab}$ that we are interested in for characterizing NUT charges. For each $B^{(1)}_{(\mu)ab}$
\be
B^{(1)}_{(\mu)ab} = \DD_a \DD_b \hat \zeta_{(\mu)} + h^{(0)}_{ab} \hat \zeta_{(\mu)},
\ee
one can integrate equation \eqref{dBk} (say, using suitable ansatz for the form of $k_{ab}$) to find a singular representative for the tensor potential $k_{ab}$. Note that, this tensor potential is defined only up to the freedom
\be
k_{ab} \rightarrow k_{ab} + \DD_a \DD_b \omega + h^{(0)}_{ab} \omega.
\ee
This freedom exists because the combination $\DD_a \DD_b \omega + h^{(0)}_{ab} \omega$ has vanishing curl, hence it does not contribute to the magnetic part of the Weyl tensor. Now recall that this freedom also exactly correspond to performing supertranslations in the space of Beig-Schmidt configurations \eqref{metric}. Since function $\omega$ depends smoothly on angular coordinates,
requiring $k_{ab}$ to contain only singular terms, for example, one can still completely fix the supertraslation ambiguity. More generally, choosing a particular representative for the inverse of the curl operator we can fix supertranslation ambiguity.

The singular representatives for $k_{ab}$ corresponding to four harmonics \eqref{zetahat} are listed in Appendix \ref{ki} in the gauge $k_a{}^{a}=0$. These representative are chosen to be regular in the north or in the south patch. The transition functions from one patch to the other take a simple form and have interesting properties. Since all four representatives have $k_a{}^{a}=0, \DD^a k_{ab} =0$, if follows from lemma \ref{BS} that for general $B^{(1)}_{ab}$, the tensor potential $k_{ab}$ can be chosen to be trace-free and divergence-free. From now on we explicitly assume
\be
k = k_{a}{}^{a}  = 0, \qquad \DD^a k_{ab} =0, \label{bc}
\ee
and moreover regard \eqref{bc} as part of our boundary conditions. Note that this boundary condition does not imply any restriction on the form of $B^{(1)}_{ab}$.

Integrating respectively electric and magnetic parts of the Weyl tensors on a cut of the hyperboloid against the four conformal Killing vectors of $dS_3$, one can define the electric and the magnetic four-momenta.

\section{Mann-Marolf Action Principle and Taub-NUT}
\label{sec:MM}
In this section we study the Mann-Marolf action principle \cite{MM} for asymptotically flat spacetimes with $k_{ab} \neq 0$. In section \ref{onshell} we show that the on-shell action is finite. We also calculate the on-shell value of the action. In section \ref{variation} we study the first variations of the action and show that it gives a well defined variational principle even when $k_{ab}$ is allowed to vary within our boundary conditions. As mentioned in the introduction, in showing this we perform integrations
by parts which are not completely justified when singularities are present in $k_{ab}$. The investigation in that aspect should
be regarded as preliminary. The main point we want to emphasize in the following is that Taub-NUT metric satisfy all
boundary conditions required to be considered in the action principle of \cite{MM}.

\subsection{On-shell Action is Finite}
\label{onshell}
We start by recalling the action principle of \cite{MM}. There it was shown that a good variational principle for asymptotically flat configurations defined by the expansion \eqref{metric} is given by the action
\begin{equation}
\label{covaction} S = \frac{1}{16\pi G} \int_{\cal M} d^4x \sqrt{-g}\,R + \frac{1}{8\pi G} \int_{\partial {\cal M}} d^3x\sqrt{-h} \,(K - \hat K),
\end{equation}
where $\hat K := h^{ab} \hat K_{ab}$ and $\hat K_{ab}$ is defined to satisfy
\begin{equation}
\label{defhatK}
{\cal R}_{ab} = \hat K_{ab} \hat K - \hat K_a{}^{c} \hat K_{cb},
\end{equation}
and where ${\cal R}_{ab} $ is the Ricci tensor of the boundary metric $h_{ab}$. Further details can be found in \cite{MM, MMV, MMMV}.

To see that the on-shell action is finite, simply note that since our spacetimes are Ricci flat, the bulk term vanishes on-shell. As a result,
\be
(16 \pi G) S_{\rom{on-shell}} = 2 \int_{\partial \cM} d^3x\sqrt{-h} (K - \hat K).
\ee
Using expansions of $K_{ab}$ \eqref{Kab} and $\hat K_{ab}$ \eqref{hatKab} from appendix \ref{app:calc} we find that
\be
(16 \pi G) S_{\rom{on-shell}} = \frac{1}{2} \int_{dS_3}d^3x\sqrt{-h^{(0)}} \left[3 \s^2 + \s_{ab} k^{ab} - \s_{ab}\s^{ab} \right]. \label{appBcalc}
\ee
In particular, note that all divergent terms have canceled, and the on-shell action is finite. Performing integrations by parts and using equations of motion for $\s$ and $k_{ab}$ we see that the on-shell action is in fact identically zero
\be
S_{\rom{on-shell}} = 0.
\ee
The interpretation of this second result is as follows.  We show in the following subsection that $\delta S = 0$ on \emph{all} variations satisfying our definition of asymptotic flatness. It then follows that $S_{\rom{on-shell}}$ must be constant as we travel along any smooth
one-parameter family of solutions. We expect any solution to be
smoothly connected to Minkowski space, where $S_{\rom{on-shell}}$ is identically zero. It then follows that $S_{\rom{on-shell}}$ is identically zero on any asymptotically flat solution.  Note that in this calculation and the argument, we have ignored the boundary contributions coming from the future or past boundaries and also from possible inner boundaries.

\subsection{Gravitational Magnetic Charges can be Varied}
\label{variation}
Now we consider the first variation of the Mann-Marolf action over configurations satisfying our boundary conditions and evaluate it on-shell. We show that the first variation vanishes about any asymptotically flat solution. Here we establish a more general result than what was shown in \cite{MM}. We show that for arbitrary variation
\be
\delta h^{(1)}_{ab} = - 2 \delta \sigma h^{(0)}_{ab} + \delta k_{ab},
\ee
the variational principle is well-defined; in particular, we will \emph{not} require as in \cite{MM} that
$\delta h^{(1)}_{ab} = -2 \delta \s h^{(0)}_{ab}$. This allows us to consistently vary gravitational magnetic charges in the variational principle. As we will see below the boundary condition \eqref{bc} plays a crucial role in this argument.

Let us start by recalling certain basic facts about the Einstein-Hilbert action with the Gibbons-Hawking term. The variation of these terms is simply
\be
(16 \pi G) \delta (S_{\rom{bulk}} + S_{\rom{GH}}) = \int_{\partial \cM} \sqrt{-h}d^3 x \pi^{ab}\delta h_{ab}, \label{EHGH}
\ee
where $\pi^{ab} = K h^{ab}- K^{ab}$ and $h_{ab}$ is the induced metric on the boundary $\partial \cM$. As can be readily checked, when the boundary $\partial \cM$ is taken to spatial infinity, the right hand side of \eqref{EHGH} does not vanish under asymptotically flat boundary conditions---the action $S_{\rom{bulk}} + S_{\rom{GH}}$ is not stationary under full class of variations preserving asymptotically flat boundary conditions.  In this sense, the Einstein-Hilbert action with the Gibbons-Hawking term added, does not provide a well defined variational principle for asymptotically flat gravity. Reference \cite{MM} showed that adding the counterterm $S_\rom{counter} = - \frac{1}{8 \pi G} \int_{\partial \cM} d^3 \sqrt{-h} \hat K$ remedies this situation. The counterterm $\hat K$ is defined via \eqref{defhatK}.

To see this we need to consider the first variation of the counterterm. This computation was done in Appendix A of \cite{MMMV}. We find
\be
(16 \pi G) \delta S_{\rom{total}} = \int_{\partial \cM} \sqrt{-h} d^3 x(\pi^{ab} - \hat \pi^{ab} + \Delta^{ab}) \delta h_{ab},
\ee
where $\hat \pi^{ab} = \hat K h^{ab}- \hat K^{ab}$ and $\Delta_{ab}$ denotes a number of extra terms. These extra terms take the form:
\be
\Delta^{ab} = \hat K^{ab} - 2 \tilde L^{cd} (\hat K_{cd} \hat K^{ab} - \hat K^{a}_{c} \hat K_{d}^{b}) + D^2 \tilde L^{ab} + h^{ab}D_c D_d \tilde L^{cd} - 2 D_{d}D^{(a}\tilde L^{b)d}.
\label{Delta}
\ee
The symbols $L_{ab}{}^{cd}$ and $\tilde L^{ab}$ are defined as follows
\be
L_{ab}{}^{cd} = h^{cd}\hat K_{ab} + \delta_{(a}^c\delta_{b)}^d \hat K - \delta_{(a}^c\hat K_{b)}^d- \delta_{(a}^d\hat K_{b)}^c,
\ee
\be
\tilde L^{ab} := h^{cd}(L^{-1})_{cd}{}^{ab}.
\ee
Following  \cite{RossLeWitt} the four index tensor $L_{ab}{}^{cd}$ is taken to be symmetric in its upper as well as lower two indices. Similarly, $\tilde L^{ab}$ is symmetric in its two indices. $D$ denotes the covariant derivative compatible with the full metric on the hyperboloid $h_{ab}$.

A simple calculation using asymptotic expansion shows that
\be
(\pi^{ab} - \hat \pi^{ab} + \Delta^{ab}) = \frac{1}{\rho^{4}}\left(\s^{ab} + \s h^{(0)\:ab}\right) + \mathcal{O}\left(\frac{1}{\rho^{5}}\right).
\label{inter}
\ee
Certain details on this computation can be found in Appendix \ref{app:calc}. Now, using $\delta h_{ab} = \rho \delta h^{(1)}_{ab} + \ldots = - 2 \rho \delta \s h^{(0)}_{ab} + \rho \delta k_{ab} + \ldots$ and $\sqrt{-h} = \rho^{3}\sqrt{-h^{(0)}} + \ldots $ it follows that in the $\rho \to \infty $ limit
\be
(16 \pi G) \delta S_{\rom{total}} = \int_{dS_3} \sqrt{-h^{(0)}} d^3 x\left( \s^{ab} + h^{(0)\:ab}\s\right)\delta k_{ab}.
\ee
In arriving at this simplified expression we have used the equation of motion for $\sigma$. Using integration by parts we then have
\be
(16 \pi G) \delta S_{\rom{total}} = \int_{dS_3} \sqrt{-h^{(0)}} d^3 x\left( -\s^{a}\delta (\DD^b k_{ab}) + \delta (h^{(0)\:ab}k_{ab})\s\right).
\ee
Now recall that we are only interested in variations over $k_{ab}$ which all satisfy the boundary condition \eqref{bc}: $\DD^b k_{ab} = 0 $ and $k_{a}{}^{a} = 0$. Therefore, we conclude that
\be
\delta S_{\rom{total}} = 0.
\ee
In particular, as shown in the previous section singular representatives for Taub-NUT (and boosted Taub-NUT) also satisfy $\DD^b k_{ab} = 0 $ and $k_{a}{}^{a} = 0$. Hence, it follows that Taub-NUT metric satisfies all
boundary conditions required to be considered in the action principle of Mann and Marolf.

One might expect that the conserved quantities now follow by a relatively straightforward boundary stress tensor construction.
 This construction is particularly desirable, since a satisfactory description of Lorentz charges in the presence of gravitational magnetic charges is not presently available in the literature.
In fact, this was one of our prime motivations for this work. However, preliminary investigation reveal that straightforward extension of \cite{MM, MMV, MMMV} lead to some interesting subtleties due of the presence of $k_{ab}$. The situation is actually worse because of the presence of singularities in $k_{ab}$. We therefore do not attempt to construct the boundary stress tensor in this paper. We will return to this problem elsewhere. 

Instead, to gain certain intuition for the nature of the Lorentz charges in the presence of gravitational magnetic charges, we express the Kerr Taub-NUT solution to the required order in the Beig-Schmidt form in the next section.

\section{Kerr Taub-NUT in Beig-Schmidt Coordinates}
\label{app:KerrNut}
In this section we express the Kerr Taub-NUT solution to the required order in the Beig-Schmidt form; thereby making contact with the singular tensor representatives presented in Appendix \ref{ki} for the first order fields, and certain other tensor harmonics relevant for the second order fields. The Kerr-Taub-NUT solution in Boyer-Lindquist coordinates is
\bea
ds^2 &=& - \frac{1}{\Sigma} \left( \Delta - a^2 \sin^2 \theta \right) dt^2 + \frac{\Sigma}{\Delta} dr^2 + \Sigma d\theta^2 + \frac{1}{\Sigma} \left( (\Sigma - a \chi)^2 \sin^2 \theta - \chi^2  \Delta \right) d\phi^2 \cr
 & & + \frac{2}{\Sigma} \left( \chi \Delta + a (\Sigma - a \chi) \sin^2 \theta \right) dt d\phi, \label{kerrTN}
\eea
with
\bea
\Sigma &=& r^2  + (N - a \cos \theta)^2, \\
\Delta &=& r^2 - 2 M r - N^2 + a^2, \\
\chi &=& - a \sin^2 \theta + 2 N (c-\cos \theta).
\eea
It is easy to check that the metric reduces to that of the Kerr black hole  upon setting $N=0$ and to Taub-NUT upon setting $a=0$.

Metric \eqref{kerrTN} has four independent parameters: $M, N, a, c$. The parameters $M$ and $N$ are interpreted as electric and magnetic masses, respectively. The parameter $a$ is the rotation parameter.
 Metric \eqref{kerrTN} is singular on the $z$-axis
as can be seen by calculating, say, $\nabla_\mu t \nabla^\mu t$. The singularity is either a singularity of the coordinates or that of the manifold.
The parameter $c$ refers to the location of the singularity. For $c=1$ the singularity is at $\theta = \pi$, for $c=-1$ it is at $\theta = 0$, for all other values of $c$ both singularities exist. For $c=0$ the north and south poles play a symmetrical role.  It was shown by Misner \cite{Misner} that this singularity is a coordinate singularity and metric \eqref{kerrTN} describes a smooth manifold provided the time coordinate is taken to be periodic with period $8 \pi N$. However, in this work, we do not view Kerr-Taub-NUT spacetime with a periodic time coordinate. Since we are interested in issues related to asymptotic flatness, we view Kerr-Taub-NUT spacetime as a singular perturbation over Minskowski space. This interpretation is justified asymptotically. It is motivated by the fact that Kerr-Taub-NUT metric is asymptotically flat at spatial infinity  \cite{Bunster:2006rt} in the sense of Regge and Teitelboim \cite{Regge:1974zd} for $c=0$.

We now show that metric \eqref{kerrTN} is also asymptotically flat at spatial infinity in the sense of Beig and Schmidt. We establish this by constructing an appropriate coordinate chart. In particular, we explicitly find a change of coordinates so that the line element is brought to the form
\bea
\label{metric2}
ds^2 &=& (( 1 + \rho^{-1}\s )^2 + \mathcal{O}(\rho^{-3}) )d \rho^2  + \rho^2\left( h^{(0)}_{ab} + \rho^{-1} h^{(1)}_{ab} + \rho^{-2} h^{(2)}_{ab} + \mathcal{O}(\rho^{-3}) \right) dx^a dx^b \nn \\
&& \: + 2 \rho (\mathcal{O}(\rho^{-3})) d \rho d x^a,
\eea
where $h^{(0)}_{ab}$ is the metric on the unit hyperboloid. It then follows from the general theorem of Beig and Schmidt \cite{BS} that appropriate coordinate transformations exist that bring the line element to the form \eqref{metric}. Reference \cite{Ashtekar:1991vb} explicitly brought Schwarzschild metric to the form \eqref{metric2}, and reference \cite{MMMV} brought Kerr metric to the form \eqref{metric2}.

We proceed by making a series of coordinate changes. The first coordinate redefinition is the one used to put  Minkowski metric in the Beig-Schmidt form
\be
r = \rho_0 \cosh \tau_0, \qquad t = \rho_0 \sinh \tau_0.
\ee
Spatial infinity corresponds to $\rho_0 \to \infty$. Expanding in powers of $\rho_0$, we find that the boundary metric is that of the unit hyperboloid at the leading order, but $ g^{(1)}_{\rho_0 \tau_0}$ and $ g^{(2)}_{\rho_0 \tau_0}$ are non-zero, and $g_{\rho_0 \rho_0}$ component of the metric is not of the form $(1 + \rho_0^{-1}\s)^2$. The next coordinate change we do is
\bea
\rho_0 &=& \rho_1 + F(\tau_1), \\
\sinh \tau_0 &=& \sinh \tau_1 - \rho^{-1}_1 (\partial_{\tau_1} F(\tau_1) + 4 M \sinh \tau_1) \cosh \tau_1,  \\
\phi_0 &=& \phi_1 + 2 \rho_1^{-1} N \csc \theta  (\cot \theta -c \csc \theta ) \sech \tau_1 \tanh \tau_1,
\eea
where
\be
F(\tau_1) = - M (\cosh \tau_1 + 2 \tau_1 \sinh \tau_1).
\ee
This coordinate change is systematically found following the algorithm outlined in \cite{BS}.  Here $F(\tau_1)$ is a supertranslation. After this coordinate change,  the metric is in the Beig-Schmidt form at first order and $g_{\rho_1 x^a} = \mathcal{O}(\rho_1^{-1})$. We can read off $\sigma$ and $h^{(1)}_{ab}$ to be
\bea
\s &=& M \cosh 2 \tau_1 \sech \tau_1, \\
h^{(1)}_{ab} dx^a dx^b  &=& 
-2 \s h^{(0)}_{ab}(\tau_1, \theta) dx^a dx^b  - 2 N k_{(0)ab}(\tau_1, \theta) dx^a dx^b,
\eea
where $k_{(0)ab}$ is given in \eqref{kiab}. The supertranslation $F(\tau_1)$ was fixed to make sure $h^{(0)ab}h^{(1)}_{ab} = - 6 \s$.

The next set of coordinate changes are more complicated than the previous two. We do not present all explicit functions. We first do
\bea
\rho_1 &=& \rho_2 + \rho_2^{-1} \tilde F(\tau_1, \theta),
\eea
where $\tilde F(\tau_1, \theta)$ is chosen so that $g_{\rho_2 \rho_2}$ component of the metric is of the form $(1 + \rho_2^{-1}\s)^2$. We then define
\bea
\sinh \tau_1 &=& \sinh \tau_2 + \rho^{-2}_2 F_1(\tau_2, \theta_2),  \\
\theta &=& \theta_2 + \rho^{-2}_2 F_2(\tau_2, \theta_2),\\
\phi_1 &=& \phi_2 + \rho^{-2}_2 F_3(\tau_2, \theta_2),
\eea
where the functions $F_1,F_2,F_3$ are found by solving the linear equations $g_{\rho_2 x^a} = \mathcal{O}(\rho_2^{-2})$.  We now rename the coordinates as  $\tau_2 = \tau, \theta_2 = \theta, \phi_2 = \phi$.  The line element is then in the form \eqref{metric2}.

The explicit expressions for the metric components for $h^{(2)}_{ab}$ are fairly complicated. We do not present them.  We have  checked using these expressions that all equations of motion are satisfied up to second order in the asymptotic expansion. At the linearized level, i.e., ignoring $M^2$ and $N^2$ terms, we have
\bea
h^{(2)}_{\tau \tau} =  - 6 Na \cos \theta \sech^4 \tau, & &
h^{(2)}_{\tau \theta} = 3 N a \sech^2 \tau \sin\theta \tanh \tau,\\
h^{(2)}_{\tau \phi} =  -3  M a \sech^2\tau  \sin ^2\theta,&&
h^{(2)}_{\theta \theta} = - 3 N a \cos \theta \sech^2 \tau,\\
h^{(2)}_{\theta \phi} = 0, &&
h^{(2)}_{\phi \phi} =  h^{(2)}_{\theta \theta}\sin^2 \theta.
\eea
In terms of the tensors $V_{(i)ab}$ and $W_{(i)ab}$  of Lemma 3 of Appendix A of \cite{CDV}, these components can be rewritten as
\be
h^{(2)}_{ab} = - 3 M a W_{(1)ab} - 3 N a V_{(1)ab} + \mbox{non-linear terms}. \label{linh2}
\ee
It follows that
\be
-\curl h^{(2)}_{ab} := -\epsilon_{a}{}^{cd}D_{c}h^{(2)}_{db} =  3 M a V_{(1)ab} - 3 N a W_{(1)ab} +  \mbox{non-linear terms}. \label{curllinh2}
\ee
Equations \eqref{linh2} and \eqref{curllinh2} call for an interpretation. These equations suggest that angular momentum of the Kerr Taub-NUT spacetime is $J = M a$ and, also, the boost charge associated to the Killing vector dual to the rotational Killing vector $\partial_\phi$, i.e., $\cos \theta \partial_\tau - \tanh \tau \sin \theta \partial_\theta$, is $K = N a$. This interpretation, even if somewhat exotic, seems to be coming out of the above equations (again ignoring singularities). The  exoticness lies in the fact that generally we do not use boost charges to label physical states. In the absence of a satisfactory construction of Lorentz charges when gravitational magnetic charges are present, we warn the reader that the above assignment of charges should not be taken too seriously, but only as a hint.

\section{Conclusions}
\label{conclu}
To summarize, we have shown that competing histories in the Mann-Marolf variational principle
can have different electric as well as magnetic four-momenta (modulo issues related to the boundary contributions due to the presence of singularities). This generalizes the previous result \cite{MM} where with hyperbolic temporal and spatial cut-offs variations over asymptotically flat configurations with only fixed magnetic four-momentum, i.e., $\delta k_{ab} = 0$, were considered.

A natural extension of this work would be to construct the boundary stress tensor at the next to leading order in the asymptotic expansion. This could then be used to construct Lorentz charges in the presence of gravitational magnetic charges. A satisfactory construction of Lorentz charges in the presence of gravitational magnetic charges is not presently available in the literature. Some comments appear in \cite{MM}. Preliminary investigation reveal that straightforward extension of the construction of \cite{MM, MMV, MMMV} lead to some interesting subtleties due of the presence of $k_{ab}$. The situation is actually worse because of the presence of singularities in $k_{ab}$. In regard to the singularities, ideas on harmonic superspace as mentioned in \cite{BMS} can be of help. Perhaps a less ambitious, though related, goal can be to study the role of supertranslation ambiguities  in the  construction of \cite{MM, MMV, MMMV} for defining Lorentz charges, i.e., taking $k_{ab}$ to be regular as well as curl-free and constructing the boundary stress-tensor and hence Lorentz charges. Investigations along these lines can teach us some important lessons about gravitational duality and/or holography in flat spacetimes. We will return to these problems elsewhere.

\subsection*{Acknowledgements}
I am indebted to Glenn Barnich, David Berman, Fran\c{c}ois Dehouck, Simon Ross and especially Geoffrey Comp\`ere and Donald Marolf for discussions. I also thank Donald Marolf and Axel Kleinschmidt for reading an earlier draft of this work.

\appendix

\section{Singular Representatives of Tensor Potential}
\label{ki}
This appendix is done in collaboration with Geoffrey Comp\`ere.  See also appendix D of \cite{CD}. The four independent non-trivial representatives $k_{(\mu)ab}$ for $\mu=0,1,2,3$ of tensor potential $k_{ab}$ are as follows:
\begin{eqnarray}
k_{(0)ab} &=& \left( \begin{array}{ccc}  0&0 & 2 \frac{c-\cos\theta}{\cosh\tau}\\ 0 &0 & \sinh\tau \frac{\cos{2\theta}-4c \cos\theta +3}{2\sin\theta} \\ 2 \frac{c-\cos\theta}{\cosh\tau} &  \sinh\tau \frac{\cos{2\theta}-4c \cos\theta +3}{2\sin\theta}&0 \end{array}\right),\nonumber \\
k_{(1)ab} &= &\left( \begin{array}{ccc}  0&0 & -3 \frac{\tanh\tau}{\cosh\tau}\sin^2\theta\\ 0 &0 & \frac{-8c+9\cos\theta - \cos 3\theta}{4\sin\theta} \cosh\tau\\ -3 \frac{\tanh\tau}{\cosh\tau}\sin^2\theta &  \frac{-8c+9\cos\theta - \cos 3\theta}{4\sin\theta} \cosh\tau  &0 \end{array}\right),\label{kiab} \\
k_{(2)ab} &= &\left( \begin{array}{ccc} 0& 3 \frac{\tanh\tau}{\cosh\tau}\sin\phi   & 3 \frac{\tanh\tau}{\cosh\tau}\cos\theta \sin\theta \cos\phi \\ 3 \frac{\tanh\tau}{\cosh\tau}\sin\phi & \frac{8c-9\cos\theta+\cos 3\theta}{2\sin^3\theta } \cosh\tau \sin\phi & \frac{\cos^4\theta-4c \cos\theta +3}{\sin^2\theta}\cosh\tau\cos\phi \\ 3 \frac{\tanh\tau}{\cosh\tau}\cos\theta \sin\theta \cos\phi &   \frac{\cos^4\theta-4c \cos\theta +3}{\sin^2\theta}\cosh\tau\cos\phi & \frac{-8c+9\cos\theta-\cos 3 \theta}{2\sin\theta}\cosh\tau \sin\phi \end{array}\right),\nonumber \\
k_{(3)ab} &= &\left( \begin{array}{ccc} 0& -3 \frac{\tanh\tau}{\cosh\tau}\cos\phi   & 3 \frac{\tanh\tau}{\cosh\tau}\cos\theta \sin\theta \sin\phi \\ -3 \frac{\tanh\tau}{\cosh\tau}\cos\phi & \frac{-8c+9\cos\theta-\cos 3\theta}{2\sin^3\theta } \cosh\tau \cos\phi & \frac{\cos^4\theta-4c \cos\theta +3}{\sin^2\theta}\cosh\tau\sin\phi \\ 3 \frac{\tanh\tau}{\cosh\tau}\cos\theta \sin\theta \sin\phi &   \frac{\cos^4\theta-4c \cos\theta +3}{\sin^2\theta}\cosh\tau\sin\phi & \frac{8c-9\cos\theta+\cos 3 \theta}{2\sin\theta}\cosh\tau \cos\phi \end{array}\right).\nonumber
\end{eqnarray}
These tensors are trace-free and divergence free and obey the equation
$(\square -3)k_{ab} = 0 $ outside of singularities at $\theta = 0$ and $\theta = \pi$. They are regular in the north patch upon choosing $c = +1$ and in the south patch upon choosing $c = -1$.
The singular transition function between the south and north patches can be written as
\bea
\delta k_{(0)ab} \equiv k_{(\mu)ab}|_\rom{South} - k_{(\mu)ab}|_\rom{North}
&=& \left( \begin{array}{ccc}  0&0 & -\frac{4}{\cosh\tau} \\ 0&0 &4\cot\theta \sinh\tau \\ -\frac{4}{\cosh\tau} & 4\cot\theta \sinh\tau & 0 \end{array}\right),\nonumber \\
\delta k_{(1)ab} \equiv k_{(1)ab}|_\rom{South} - k_{(1)ab}|_\rom{North}  &=& \left( \begin{array}{ccc} 0 &0 & 0\\ 0&0 & 4\frac{\cosh\tau}{\sin\theta} \\ 0& 4\frac{\cosh\tau}{\sin\theta}& 0\end{array}\right), \\
\delta k_{(2)ab} \equiv k_{(2)ab}|_\rom{South} - k_{(2)ab}|_\rom{North}  &=& \left( \begin{array}{ccc} 0 &0 &0 \\0 & -\frac{8}{\sin^3\theta}\cosh\tau \sin\phi & \frac{8 \cos\theta}{\sin^2\theta}\cosh\tau \cos\phi \\ 0&\frac{8 \cos\theta}{\sin^2\theta}\cosh\tau \cos\phi  &  \frac{8}{\sin\theta} \cosh\tau \sin\phi \end{array}\right), \nn \\
\delta k_{(3)ab} \equiv k_{(3)ab}|_\rom{South} - k_{(3)ab}|_\rom{North}  &=& \left( \begin{array}{ccc} 0 &0 &0 \\0 & \frac{8}{\sin^3\theta}\cosh\tau \cos\phi & \frac{8 \cos\theta}{\sin^2\theta}\cosh\tau \sin\phi \\ 0&\frac{8 \cos\theta}{\sin^2\theta}\cosh\tau \sin\phi  &  -\frac{8}{\sin\theta} \cosh\tau \cos\phi \end{array}\right).\nonumber
\end{eqnarray}
These transition functions obey
\begin{eqnarray}
\DD_{[a} \delta k_{(\mu) b]c} = 0, \quad (\square - 3) \delta k_{(\mu)ab} = 0,\quad  h^{(0)\:ab}\delta k_{(\mu)ab}=0,\quad \DD^b \delta k_{(\mu)ab}=0, \label{propdeltak}
\end{eqnarray}
on the hyperbolid outside of the singularities. 
 When integrated against conformal Killing vector of $dS_3$, $\DD^a \zeta_{(\nu)}$, these transition functions obey orthogonality relations
\bea
\int_0^{2\pi} d\phi \;\delta k_{(\mu) \phi a}\DD^a \zeta_{(\nu)} = -8\pi \; \eta_{(\mu)(\nu)},\qquad \mu,\nu=0,\dots 3,\label{orthok}
\eea
where $\eta_{(\mu)(\nu)}$ is a diagonal matrix with entries $(-1,1,1,1)$.

\section{Calculation of $K_{ab}$ and $\hat K_{ab}$}
\label{app:calc}
In this appendix we follow appendix B of \cite{MMV} to calculate expansions of $K_{ab}$ and $\hat K_{ab}$. We make use of the asymptotic field equations as needed.  The extrinsic curvature $K_{ab}$ expands as
\be
K_{ab} = \rho h^{(0)}_{ab} + \left[\frac{1}{2} k_{ab} - 2 \s h^{(0)}_{ab} \right] + \rho^{-1}\left[ 2 \s^2 h^{(0)}_{ab} - \frac{1}{2} \s k_{ab}\right] + \ldots,
\label{Kab}
\ee
and its trace as
\be
K = \frac{3}{\rho} + \frac{1}{\rho^3} \left[ 6 \s^2 + \frac{1}{2} k_{ab}k^{ab} - h^{(2)}\right] + \ldots~.
\label{traceK}
\ee
With these expressions at hand $\pi_{ab} = K h_{ab} - K_{ab}$ can be readily calculated.

The calculation of $\hat K_{ab}$ is a bit more involved. We first need the asymptotic equations of motion at second order.  After a tedious computation we find that on-shell
\bea
R^{(2)}_{ab} &=& 2 h^{(2)}_{ab} - 9 \s^2 h^{(0)}_{ab} - 2 \s_c \s^c h^{(0)}_{ab} + 2 \s_a \s_b - \s \s_{ab} - \s_{cd} k^{cd} h^{(0)}_{ab} \nn \\
&& - \frac{1}{2}\s^{c} \left( \DD_a k_{bc} + \DD_b k_{ac} - \DD_c k_{ab}\right) + \s k_{ab} + \frac{1}{4}k_{cd}k^{cd}h^{(0)}_{ab} - \frac{1}{2}k_a^{c} k_{cb},
\eea
and
\be
R^{(2)} := h^{(0)\:ab} R^{(2)}_{ab} =  - 2 \s_c \s^c - \s_{cd}k^{cd} + \frac{3}{4} k_{cd}k^{cd}.
\ee

Expanding $\hat K_{ab}$ as
\be
\hat K_{ab} = \rho h^{(0)}_{ab} + \hat p^{(1)}_{ab} + \rho^{-1}  \hat p^{(2)}_{ab} + \ldots,
\label{hatKab}
\ee
and inverting the defining relation \eqref{defhatK} in the power series expansion following appendix B of \cite{MMV}, and using the asymptotic equations of motion we find
\be
\hat p^{(1)}_{ab} = \s_{ab} - \s h^{(0)}_{ab} + \frac{1}{2}k_{ab},
\ee
and
\bea
\hat p^{(2)}_{ab} &=& h^{(2)}_{ab} - \s_{c}\s^c h^{(0)}_{ab}  + \s \s_{ab} + 2 \s_a \s_b + \s_{ac}\s_b{}^{c} - \frac{1}{4} \s^{cd}\s_{cd}  h^{(0)}_{ab} - \frac{5}{4} \s^2h^{(0)}_{ab} \nonumber \\
& & - \s_{c(a}k_{b)}{}^{c} - \frac{1}{2}\s^c (\DD_{b}k_{ac} + \DD_{a}k_{bc}- \DD_{c}k_{ab}) + \frac{1}{4}\s_{cd}k^{cd}h^{(0)}_{ab} - \frac{1}{4} k_{a}{}^{c} k_{cb}.
\eea
Taking the trace of these expressions gives
\be
\hat p^{(1)} = - 6 \s, \qquad \hat  p^{(2)} = \frac{21}{4}\s^2 + \frac{1}{4} \s_{cd}\s^{cd} + \frac{3}{4} \s_{cd}k^{cd}.
\label{tracehatK}
\ee
Putting various pieces together we obtain equation \eqref{appBcalc}:
\bea
K - \hat K = \frac{1}{4\rho^3} \left[ 3 \s^2 - \s_{ab}\s^{ab} + \s_{ab}k^{ab}\right] + \mathcal{O}(\rho^{-4}).
\eea
Now, performing a straightforward calculation using these results and following appendix B\footnote{Appendix B of reference \cite{MMMV} is only available in the arxiv version of the paper.} of \cite{MMMV} we find that $\Delta_{ab}= \mathcal{O}(\rho^{-1})$. Equation \eqref{inter} then follows.


\end{document}